\documentclass[pra,%
 reprint,
 amsmath,amssymb,
 aps,superscriptaddress,
]{revtex4-2}
\usepackage{lipsum}
\usepackage{braket}
\usepackage{booktabs}
\usepackage{graphicx}% Include figure files
\usepackage{dcolumn}% Align table columns on decimal point
\usepackage{bm}% bold math
\usepackage{hyperref}% add hypertext capabilities
\usepackage{setspace}
\usepackage[normalem]{ulem}
\usepackage{color}
\usepackage[dvipsnames]{xcolor}

\begin{document}

\preprint{APS/123-QED}

\title{ Entanglement estimation of Werner states with a quantum extreme learning machine }
%\title{Manuscript Title:\\with Forced Linebreak}% Force line breaks with \\
\author{Hajar Assil}
\affiliation{Laboratory of R\&D in Engineering Sciences, Faculty of Sciences and Techniques Al-Hoceima, Abdelmalek Essaadi University, Tetouan, Morocco}

\author{Abderrahim El Allati}
\affiliation{Laboratory of R\&D in Engineering Sciences, Faculty of Sciences and Techniques Al-Hoceima, Abdelmalek Essaadi University, Tetouan, Morocco}
\affiliation{Max Planck Institute for the Physics of Complex Systems, Nöthnitzer Str. 38, D-01187 Dresden, Germany}

\author{Gian Luca Giorgi}%
\affiliation{%
 Institute for Cross-Disciplinary Physics and Complex Systems (IFISC) UIB-CSIC, Campus Universitat Illes Balears, 07122 Palma de Mallorca, Spain.
}%
\email{gianluca@ifisc.uib-csic.es}

\begin{abstract}
Quantum Extreme Learning Machines (QELMs) have emerged as a potent tool for various quantum information processing tasks. We present a QELM protocol for estimating the amount of entanglement in Werner states. 
The protocol requires the generation of a sequence of random Werner states, which are then combined with a reservoir state and evolved using an Ising Hamiltonian. A set of observables based on the Bloch basis is constructed and employed to train the system to recognize unseen features.
To assess the protocol's robustness, noise is introduced into the input states, and the system's performance under these noisy conditions is analyzed. Additionally, the influence of the magnetic field parameter within the Ising Hamiltonian on the estimation accuracy is investigated.

\end{abstract}

\maketitle

%\tableof

\section{Introduction}

Traditional machine learning algorithms face challenges as datasets become more complex and larger. Quantum computing presents a promising solution to these limitations with its inherent parallelism and exponential computational capabilities. By utilizing the distinct characteristics of quantum systems, Quantum  Machine learning can reach new frontiers in speed, precision, and scalability for training neural networks \cite{wittek2014quantum,schuld2015introduction,biamonte2017quantum,dunjko2018machine}.

A very successful machine learning paradigm is represented by reservoir computers (RCs) and extreme learning machines (ELMs), which are subclasses of recurrent neural networks (RNN) that do not require the tuning of the parameters of the reservoir, as during the learning process, only the weights of the output layer are adapted through linear regression \cite{ESM,maass2004computational,nakajima2021reservoir,huang2011extreme}. Besides the easiness of the training process,  reservoir computers and extreme learning machines are known for their ability to solve multiple tasks simultaneously by training the output layer independently for each task. The primary distinguishing factor between RCs and ELMs is the reservoir's internal memory utilization. While reservoir computers retain the memory of past inputs, making them appropriate for processing temporal data, extreme learning machines perform static tasks and are not required to possess internal memory. %The quantum counterparts of both RCs and ELMs.

The extension to the quantum domain of the RC \cite{fujii2017harnessing,Mujal2021,Nokkala2021,martinez2021dynamical,PRXQuantum.3.030325,kornjavca2024large,senanian2024microwave}
  and ELM \cite{QuantumInfoProcessing,ghosh2019quantum,Innocenti2023,  PhysRevLett.132.160802,xiong2023fundamental} paradigms are motivated by the promise of the exponential size of the Hilbert space compared to the physical size of the system \cite{Martinez2020,PhysRevResearch.4.033007,Dudas2023, sakurai_2022_QRCscale_free}, which can potentially lead to overcome classical algorithms, and with the intrinsic ability to process quantum inputs \cite{nokkala2023online,nokkala2024retrieving} without requiring any information pre-processing, which would be inefficient in classical hardware. Furthermore, the popularity of QRC and QELM is motivated by their immediate applicability in noisy intermediate-scale quantum (NISQ) devices \cite{preskill2018quantum,bharti2022noisy}. 

%The Quantum Extreme Learning Machine (QELM) is a quantum computing approach inspired by its classical counterpart, the Extreme Learning Machine (ELM). Moreover, it is a machine learning paradigm that exploits the natural dynamics of an input-driven, randomly connected Neural Network for information processing \citep{huang2011extreme,lukovsevivcius2012practical}. In general, ELM is made up of three layers: an input layer, a reservoir that remains unchanged over time and plays a central role in processing information, and an output layer. 

The concept of ELM was introduced within the broader framework of reservoir computing, which evolved from echo state networks (ESN) and liquid state machines (LSM) in the early 2000s \citep{hauser2021physical}, \citep{Martinez2020}. The RC system consists of a reservoir denoted as $\mathcal{R}$ and a trainable readout layer $w$. The reservoir retains its shape over time and is instrumental in processing information. Various reservoir types, such as traditional RNNs, ESNs, LSMs, and physical systems like octopus arms or photonic reservoirs, can be employed. The reservoir serves as an information transducer, transforming input information into the system's state. Notably, from ELM, the state $\mathbf{x}_k$ of the reservoir is uniquely determined by the corresponding input $u_k$, as expressed by the equation 
\begin{equation}
\mathbf{x}_k=f\left(u_k\right).
\end{equation}
This design choice differs from reservoir computing, where the state evolution depends on both current and past inputs, which makes ELM useful in classification tasks. Consider a classification problem that aims to determine whether a picture contains a car or not. The presence of a car in a previous picture has no bearing on the presence of a car in the current picture\citep{Mujal2021}. 

Quantum Extreme Learning Machine (QELM) stands as an interesting form within the context of Quantum Reservoir Computing (QRC), a paradigm that seamlessly integrates quantum computing principles with physical reservoir computing for real-time machine learning applications \citep{fujii2021quantum, fujii2017harnessing}. While QRC processes input data over time, capturing temporal dependencies and correlations among previous inputs, QELM introduces a distinctive departure from this temporal paradigm. Compared to QRC, QELM does not rely on temporal dependencies and does not store information about previous inputs in the quantum reservoir. Instead, QELM performs a static, one-shot transformation of the input data using the quantum reservoir. This characteristic renders QELM particularly suitable for tasks such as classification, where the inherent temporal dynamics of sequential data may be less crucial. The novel approach of QELM capitalizes on the unique features of quantum systems, including the exponential growth of degrees of freedom in a Hilbert space (d.o.f $\sim 4^N$ for a spin-1/2 system). The temporal independence of QELM makes it an intriguing model for specific applications within the broader landscape of Quantum Reservoir Computing. 

In this paper, we investigate an extreme learning machine technique that employs a quantum network consisting of randomly coupled spins, similar to the proposals described in Fujii et al. (2017). Our goal is to accurately determine the Werner state parameter $p$ using a machine learning algorithm that relies solely on individual qubit measurements. The protocol's resilience to noise in input states is evaluated, shedding light on its efficacy under noisy conditions. 
Furthermore, we explore the concept of generalizability in our QELM. We train the network using data from various domains and then assess its performance on a target domain not included in the training set. This analysis provides insight into the system's capability to learn and apply its knowledge to unseen data, a crucial aspect for practical applications.

The structure of this paper is outlined as follows: In the initial section, we articulate our proposed model, utilizing the Transverse Ising model as a reservoir, illuminating each step in the process. Subsequently, we assess the robustness of the reservoir by introducing noise to the input states. The results section is dedicated to an in-depth discussion and analysis of the various findings.

\section{Protocol}\label{sec:model}
{\it Reservoir -}
%We used as a reservoir of our protocol the Transverse Ising Model, which is a system that can be used for a qubit implementation of QRC, with:
Our reservoir is given by a Transverse Ising Model \cite{fujii2017harnessing,Martinez2020,martinez2021dynamical}, whose Hamiltonian reads 
\begin{equation} \label{eqn:Hamiltonian}
H=\sum_{i<j}^N J_{i j} \sigma_i^x \sigma_j^x+h \sum_{i=1}^N \sigma_i^z
\end{equation}  
where $i,j$ label the sites of the network, $\sigma_i^a(a=x,y,z)$ are the Pauli matrices acting on the $i-$th site, $h$ is the value of the external magnetic field, $J_{ij}$ is the spin-spin coupling,  randomly selected from a uniform
distribution in the range $\{-J_s/2, J_s/2\}$ with $J_s = 1$  and $N$ is the number of qubits.
As discussed in \cite{martinez2021dynamical,xia2022reservoir,llodra2024} in the context of QRC, the dynamical phase of the reservoir can play an important role in the computational capabilities of the system. In particular, it has been shown that the ergodic phase, which appears for $h  \gtrsim 0.1 J_s$,  provides a suitable scenario for efficient information propagation, in contrast to the nonergodic phase, where the presence of local conserved quantities can cause slow convergence. On the other hand, such an analysis is lacking for an extreme learning machine, whose memory requirements are different from those of a reservoir computer. For the model \ref{sec:model}, the ergodic phase is achieved for magnetic fields larger than the typical values of $J_{i,j}$ (while the concept of dynamical phase transition can only strictly be applied to the case of thermodynamic systems, effects can also be seen for finite-size models, even though the transition between the two phases will not be sharp.).

{\it Werner states -}
Werner states  of two qubits are mixed states that can be defined as follows: %in this paper we will work with the case of two-qubit Werner states:
\begin{equation}
%\left|\psi_W(p)\right\rangle
\varrho%^{(2)}%
_W(p)=\frac{1-p}{4} \mathbb{I}+p\left|\psi_{-}\right\rangle\left\langle\psi_{-}\right|
\label{eq.werner}
\end{equation}
where $\mathbb{I}$ is the $4\times 4$ identity matrix and $\left|\psi_{-}\right\rangle$
% $$
% \left|\psi_{-}\right\rangle=\frac{1}{\sqrt{2}}(|01\rangle-|10\rangle)
% $$
is the singlet state. These are known to be entangled if $p>1 / 3$ and factorized otherwise \cite{PhysRevA.40.4277}.

{\it QELM protocol -}
Our goal is to estimate the parameter $p$ using a QELM, determining in such a way the entanglement of a Werner state. First, we prepare a series of  input density matrices $\varrho^{(2)}_W(p_k) =\varrho_W(p_k)$ %states 
% $\left|\psi_k\right\rangle \equiv\left|\psi_W\left(p_k\right)\right\rangle$, whose density matrix is 
% \begin{equation}
% \varrho_W\left(p_k\right)=\left|\psi_k\right\rangle\left\langle\psi_k\right|
%\end{equation}
%with $p_k\in[0,1]$.
where $p_k$ is a random number  $p_k\in[0,1]$.
Second, we prepare the density matrix of the whole system at $t=0$
\begin{equation}
\rho_k(0)=\varrho_W\left(p_k\right) \otimes R_k
\end{equation}
where $R_k$ is a random state of the reservoir selected from a uniform distribution. It is worth remarking that $R_k$ depends on $k$, that is, it is different for each input $\varrho_W\left(p_k\right)$, which makes the task highly nontrivial. Let us also remark that, while $k$-dependent, the reservoir state $R_k$ is not related to the input and does not enter the machine learning part of the protocol. 
Then we let the state evolve according to
\begin{equation}\label{eqn:rho_t}
\rho_k(t)=e^{-i H t} \rho_k(0) e^{i H t},
\end{equation}
where $e^{-i H t}$ is the operator of the unitary dynamics.
Third, the extraction of information from the density is achieved through measurement, where the corresponding observables are defined as the elements of the N-qubit Bloch vector and can be calculated as $x_i(\Delta t)=\left\langle B_i\right\rangle=\operatorname{Tr}\left[B_i \rho(\Delta t)\right]$,
where $B_i$ are the Pauli strings of the quantum reservoir and $\Delta t$ is the time at which the system is measured \cite{fujii2017harnessing}. In our paper, we will use the local observables $B_i= \sigma_i^z$. This choice is motivated by its experimental simplicity, making it practical and feasible to implement in real-world setups.  Thus, the data are collected as follows
\begin{equation}\label{eq:obs}
x_i(\Delta t)=\left\langle\sigma_i^z\right\rangle=\operatorname{Tr}\left[\sigma_i^z \rho(\Delta t)\right].
\end{equation}

Following the data acquisition from the reservoir, the subsequent step involves initiating our machine learning procedure. In this process, %we prioritize a specific set of observables, represented by $\left[\sigma_i^z \rho(\Delta t)\right]$ for each qubit. These 
observables of Eq. (\ref{eq:obs}) are then employed in training the system. To initiate the training process, we define the vector $\textbf{x}^{(k)}(\Delta t)=(x_1^{(k)}, x_2^{(k)},..., x_N^{(k)})^\top $.
In the training phase, a dataset comprising $L<n$ instances (with $n$ denoting the number of inputs) is collected, where each instance is paired with a target value $\bar{p}_k$. The objective during training is the minimization of the mean square error (MSE) between the estimated parameter $p_k$ and its corresponding target $\bar{p}_k$:
\begin{equation}
M S E^{\text {training }}=\frac{1}{L} \sum_{k=1}^L\left(p_k-\bar{p}_k\right)^2 
\end{equation}
The estimation of $p_k$ involves a weighted sum of observables, with weights $w_i^{\mathrm{LR}}$ determined using the Moore-Penrose pseudo-inverse 
$X^{+}=\left(X^T X\right)^{-1} X^T$, where $X$ represents the matrix of input data. Specifically, the weights are calculated as $\mathbf{w}^{\mathrm{LR}}=X^{+} \mathbf{y}$, where $\mathbf{y}$ is the vector of target values. Thus, the parameter $p_k$ is estimated as:

\begin{equation*}
p_k=\sum_{i=0}^N w_i^{\mathrm{LR}} x_i^{(k)}.
\end{equation*}
After minimization, we proceed to the test phase, where performance is evaluated on a particular set of $n-L$ instances. The evaluation involves computing the mean square error (MSE) for the test set. This procedure provides a quantitative measure of the system's capability to estimate the parameter $p_k$ and, consequently, the entanglement properties of the input states. A diagrammatic description of the whole protocol described here can be found in TABLE \eqref{protocol}.

\begin{table} [h!]
\centering 
\caption{Schematic representation of how our QELM protocol processes information.}
  \begin{tabular}{c|c}
      \hline
       Input &  $\varrho_W(p)=\frac{1-p}{4} \mathbb{I}+p\left|\psi_{-}\right\rangle\left\langle\psi_{-}\right|$  \\
       \\
             &  $\varrho_W\left(p_k\right) \otimes R_k$  \\
             \\
       Reservoir & $H=\sum_{i<j}^N J_{i j} \sigma_i^x \sigma_j^x+h \sum_{i=1}^N \sigma_i^z$ \\
       \\               
                &\label{eqn:rho_t1}
    $\rho_k(t)=e^{-i H t} [\varrho_W\left(p_k\right) \otimes R_k] e^{i H t} $\\
       \\
         &    $x_i(k \Delta t)=\operatorname{Tr}\left[\sigma_i^z \rho(k\Delta t)\right]$ \\
         \\
         Output &  $p_k=\sum_{i=0}^N w_i^{\mathrm{LR}} x_i^{(k)}$\\
       \hline
   \bottomrule
\end{tabular}
\label{protocol}
\end{table}
%\subsection*{Exploring Noisy Input States}
 %In any realistic application of real-world quantum systems, one needs to take into account the presence of imperfections and the interaction with some external environment that can be responsible for various sources of noise. Here, we will consider the case of a noisy input injection, where the actual input injected into the reservoir is not the ideal one and is modified as follows:

 Real-world quantum systems inevitably exhibit imperfections. These imperfections can arise from internal limitations or interactions with the surrounding environment. This can introduce various forms of noise that can hinder the performance of quantum algorithms. 
In this work, we focus on the specific case of noisy input injection. This scenario occurs when the actual input data fed into the quantum system deviates from the ideal input due to noise. For this, let us assume that the state injected into the reservoir is
\begin{equation}
 \varrho_W\left(p_k, \varepsilon\right)=(1-\varepsilon) \varrho_W\left(p_k\right)+\varepsilon r_k \label{noise_density},
 \end{equation}
where $r_k$ is a further random density matrix and $0 \leq \varepsilon \leq 1$ is a measure of the noise. 

Moreover, we will explore the impact of the magnetic field in the Hamiltonian {\ref{eqn:Hamiltonian}} on the performance of the protocol. By varying the strength of the magnetic field $h$, we want to see how this parameter affects the resilience of the quantum reservoir to noise-induced fluctuations. 
To have a closer look at our protocol, in the next sections, we will evaluate {the MSE for the test set. }

\section{Results}
\begin{figure*}
    \centering
    \includegraphics[width=0.75\linewidth]{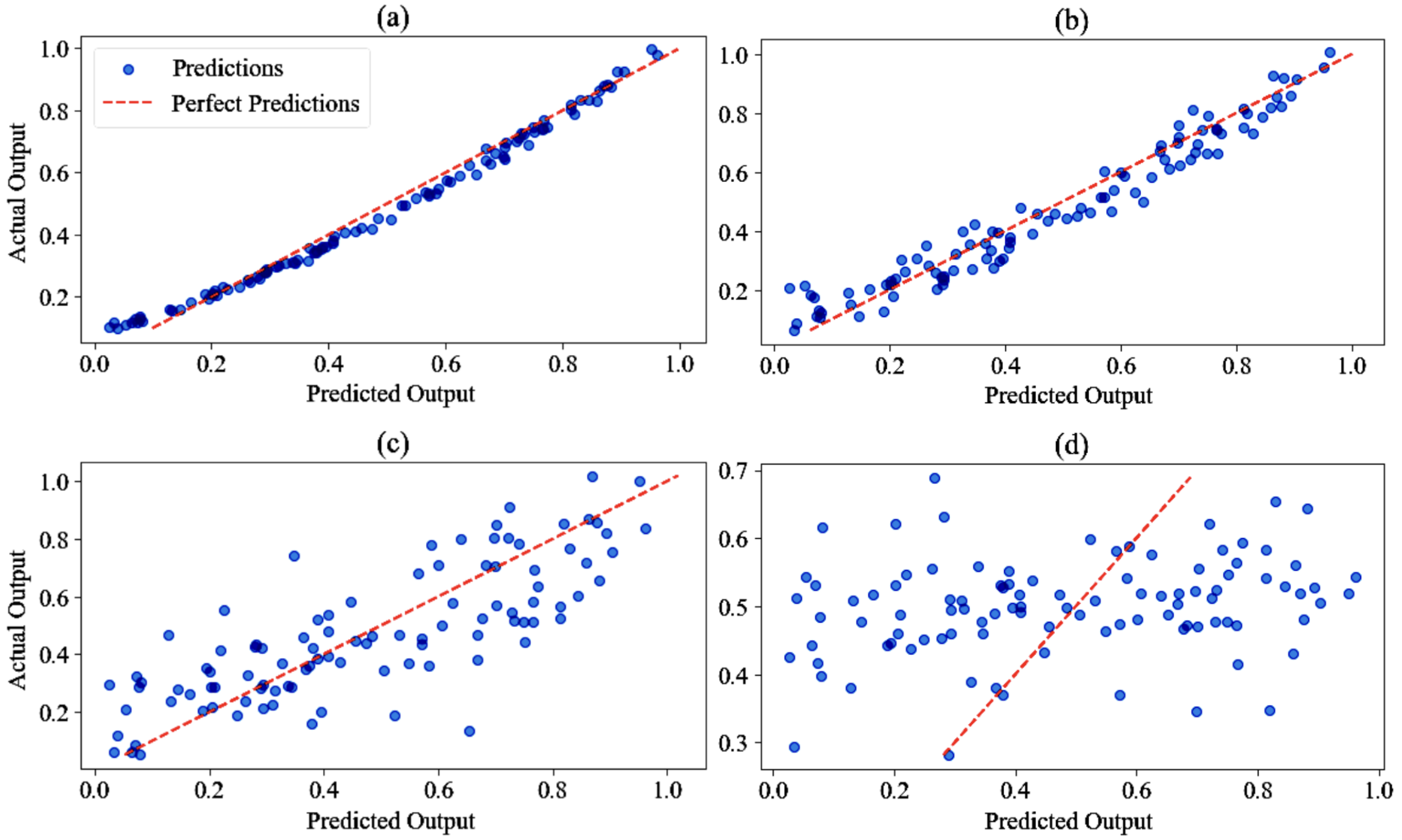}
    \caption{Comparison of Actual Output and Predicted Output in Linear Regression, (a) without noise, (b) with Low Noise $\varepsilon=0.2$, (c)  with Moderate Noise $\varepsilon=0.5$, (d) with High Noise $\varepsilon=0.9$. The figure shows the comparison between the actual output and the predicted output obtained from a linear regression model operating in the presence of varying noise levels. The blue data points represent the actual output values, while the red line represents the ideal predictions. As noise increases $(\varepsilon=0.2,  \varepsilon=0.5$, and $\varepsilon=0.9$ ), the model's ability to accurately capture the underlying trend faces challenges.}
   \label{fig:raw_data}
\end{figure*}

The protocol under consideration in this work is to estimate the parameter "$p$" for predicting the entanglement of Werner states using a system {of $N=5$ qubits. This system size will be used throughout the paper unless explicitly specified otherwise}. {For training and test, we use 100 training states and 100 test states.}
To ensure a thorough comprehension of the following study, we first present in Fig. \ref{fig:raw_data} a series of panels depicting the raw data for both the outputs and predictions.

Figure \ref{fig:raw_data} serves as a comprehensive representation of the linear regression model's performance in predicting actual data. The red line represents the target data, and the blue data points present the output of our protocol. Panels (a) through (d) describe distinct scenarios; panel (a) presents the raw data in the no-noise case, where the protocol’s output is very close to the perfect predictions, and the model demonstrates a robust ability to generalize from training data and reproduce fundamental relationships. From (b) to (d) the noise increases from $\varepsilon=0.2$ to $\varepsilon=0.9$.  Low noise scenarios show the model's resilience, maintaining accuracy despite noise (e.g., $\varepsilon=0.2$). On the other hand, high noise levels (e.g., $\varepsilon=0.9$) expose limitations, with predictions deviating significantly from actual data points. %This rigorous analysis provides valuable insights into the model's limitations and strengths, contributing to a nuanced understanding of its applicability in real world settings. 

\begin{figure}
    \centering
    \includegraphics[width=1\linewidth]{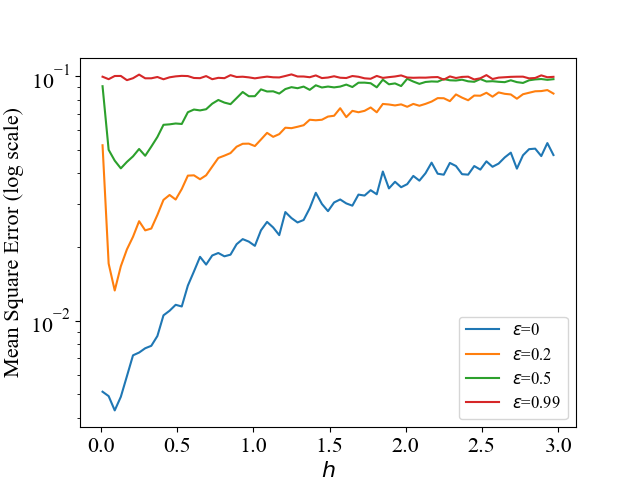}
    \caption{{MSE} as a function of the magnetic field $h$ with varying levels of noise  (Average over $20$ realizations).}
    \label{fig:correlation}
\end{figure}
The quantitative analysis starts in Fig. \ref{fig:correlation}, where we show the {MSE} as a function of the external magnetic field with different levels of noise. 
We begin by analyzing the simplest scenario: the absence of noise. This allows us to isolate and examine the core dynamics of the system. Without the confounding factor of noise, the QELM can effectively model the system using a relatively simple network configuration. This configuration does not require incorporating the magnetic field parameter $h$ into the model. For our analysis, the noise-free case serves as a baseline for evaluating the model's performance when noise is introduced.

The orange and green lines show how the dynamics of the system becomes substantially more complex when noise is introduced. It becomes evident that the model’s performance gradually diminishes. However, as noise intensifies to a considerable
degree, the model’s ability to provide accurate predictions is notably
impeded. Surprisingly, leveraging the impacts of the phase transition might be beneficial in this
case. On the contrary, there appears to be little advantage to operating
within the ergodic phase. 
The criticality of working within specific phases is
highlighted by the orange and green lines in Fig. \ref{fig:correlation}. 
These phase transitions can be strategically employed to our advantage.

The introduction of noise creates an intricate interplay of factors, and
by navigating these critical points, we gain the ability to exploit their
inherent properties to improve task performance. Intriguingly, we observe that the ergodic phase, while known for its memory-preserving
capabilities, does not necessarily align with our task’s requirements. This
insight stems from the recognition that the current task prioritizes information distribution over the retention of past inputs, rendering memory preservation less crucial. As a result, our exploration unveils the potential to optimize performance by effectively addressing noise-induced
complexities through phase transition strategies.

\begin{figure}
    \centering
    \includegraphics[width=1\linewidth]{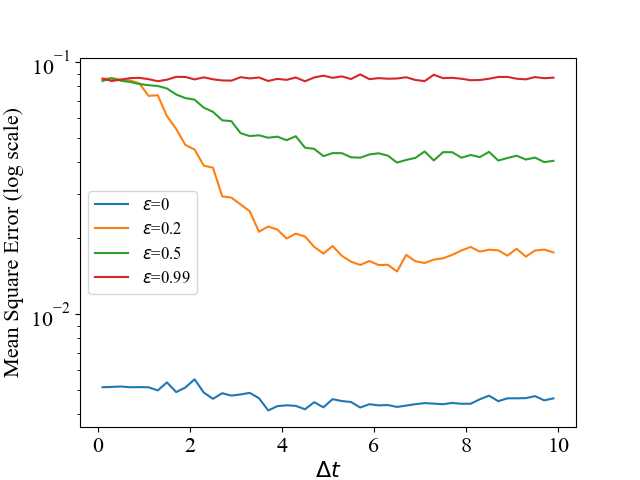}
    \caption{%Correlation between target and output in function of $\Delta t$. The system's parameter are $h=0.1$ and $N=5$ (Average of 20 realizations)
    {MSE} as a function of $\Delta t$. The system's parameter is $h=0.1$ (average of 20 realizations).  }
    \label{fig:delta_t}
\end{figure}

Figure \ref{fig:delta_t} explores how adjusting the time step size $(\Delta t)$ affects the model's ability to predict the target behavior. The experiment focuses on a system with a specific external magnetic field strength $h = 0.1$.
When there's no noise in the data, the model performs well regardless of the time step size. Any time step allows the model to capture the underlying dynamics and achieve perfect predictions.
 When the noise level is high, the model struggles to make accurate predictions irrespective of the time step. The high noise overwhelms the system, making it difficult to extract meaningful information. In the intermediate noise scenario, a minimum time step $(\Delta t)$ exists below which the model's performance suffers. This suggests that under moderate noise, the model needs a certain amount of time (captured by $(\Delta t)$) to develop the necessary internal complexity to accurately predict the target behavior.  {This can be related to the scrambling time of the reservoir \cite{vetrano2024state}.}

A relevant issue in a realistic set-up is the presence of finite resources. For instance, in any experimental realization, the observables that we introduced above can only be estimated by realizing a series of (finite) measurements (see \cite{mujal2023time,PhysRevApplied.20.014051} ) for more details. Indeed, a QRC or QELM's capability can be severely limited.   

\begin{figure}
    \centering
    \includegraphics[width=1\linewidth]{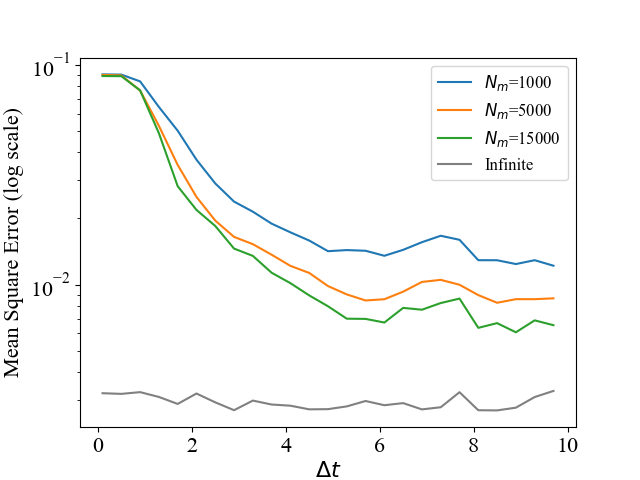}
    \caption{{MSE as a} function of $\Delta t$ with different number of measurements and infinite number of measurements $N_m$ and $\varepsilon = 0.2$. The system's parameter is $h=0.1$ (Average of 20 realizations) }
    \label{fig:delta_t_F}
\end{figure}

 Figure \ref{fig:delta_t_F} compares the results obtained previously (Fig. \ref{fig:delta_t}) using different numbers of finite measurements $N_m$ with those obtained in the infinite measurement limit. Here, the observable is evaluated for three different values of $N_m=1000,5000$, and $15000$, with the error parameter $\varepsilon= 0.2$. The {grey} line represents the results obtained in the infinite measurement limit.
In quantum mechanics, the measurement process is inherently probabilistic, and the accuracy of the measured observable depends on the number of measurements $N_m$ performed. For a finite number of measurements, the observable $\langle \mathcal{O}\rangle_{\text {finite }}$ can be expressed as:
$$\langle \mathcal{O}\rangle_{\text {finite }}=\langle \mathcal{O}\rangle_{\text {infinite}}+O\left(\frac{1}{\sqrt{N_m}}\right)$$
where $\langle \mathcal{O}\rangle_{\text {infinite}}$ is the ideal value of the observable in the infinite measurement limit, and $O\left(\frac{1}{\sqrt{N_m}}\right)$ represents the statistical error due to the finite number of measurements \cite{sannia2024dissipation, mujal2023time}.
In the figure, the solid lines correspond to different finite measurements, showing how the observable converges towards the infinite measurement limit as $N_m$ increases. We see that in our case, a relatively small number of repeated measurements is sufficient to achieve good performance in the case of limited error.

\subsection{Domain generalization}

In the context of machine learning, domain generalization refers to the problem of training a model on data from multiple source domains and then using it to perform well on a target domain that was not seen during the training stage. It is a hard problem because machine learning models are typically designed to perform well in a specific domain, and their performance may degrade when applied to new, unseen domains. 
The primary objective of domain generalization is to improve the robustness and adaptability of machine learning models to a wide range of data distributions. This is particularly critical for real-world applications where the data can vary greatly in various scenarios, making it impractical to collect labeled data for every potential domain. 

\begin{figure}[h!]\label{fig:gen}
    \centering
    \includegraphics[width=.9\linewidth]{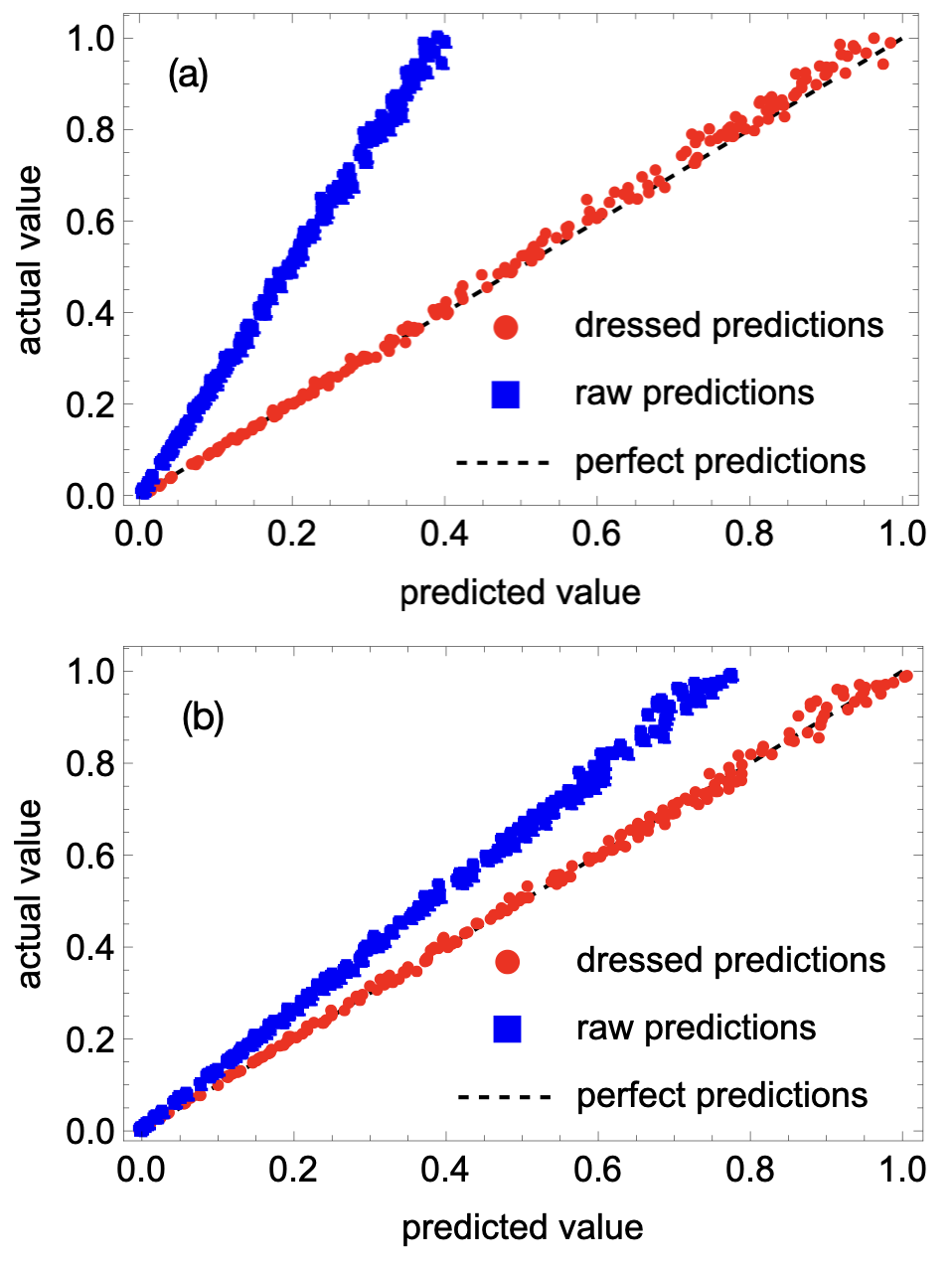}
    \caption{Ability of the QELM model to generalize to unseen data from a new domain. The blue line represents the raw data obtained from the generalized domain (different from the training data). The red line represents the ``dressed data". which is constructed by adding the input-output relationship of a single element from the target domain (the domain the model was trained on) to the raw data points in the generalized domain. }
    \label{generalization}
\end{figure}

In the following, we demonstrate the ability of the QELM to generalize the learning of Werner states in higher dimensions, which is possible with knowledge of a single element in the test domain. More specifically, during the training phase, we will utilize the Werner states defined in Eq. (\ref{eq.werner}). These states will be relabeled as $\varrho^{(2)}_W(p)$, where the superscript $2$ denotes the number of qubits utilized to encode the state. Then, in the test stage, the reservoir will be fed with fresh states belonging to  
\begin{equation}
   \varrho^{(n)}_W(p)= \frac{1-p}{2^{n}} \mathbb{I}_n+p\left|\psi_{GHZ}\right\rangle\left\langle\psi_{GHZ}\right|,
\end{equation}
where $\mathbb{I}_n$ is the $(2^n\times 2^n)$ identity matrix and $|\psi_{GHZ}\rangle$  is any GHZ state (for the sake of clarity, in Fig. \ref{fig:gen} we used $|\psi_{GHZ}\rangle=(|0,0,\dots,0\rangle+|1,1,\dots,1\rangle)/\sqrt{2}$).
This can be seen as a generalization of Werner states to larger dimensions $n>2$. To maintain the reservoir connections unchanged {in the training and in the test phase}, we will keep {the }total number of qubits { $N$ fixed}.

In Fig. \ref{generalization}, we present the prediction made for targeting Werner states of $n=3$ (panel (a)) and $n=4$ (panel (b)) qubits. In both cases, we used $N=7$. We compare the perfect prediction (diagonal dashed lines) with the actual output of the QELM (blue points). The predictions (referred to as "raw predictions" in the figure) demonstrate {that the reservoir is able to capture the existence of a linear dependence between the target and the output, even though it lacks the ability to predict the exact value of the target. In order to do that, we only need to know the target of a single element from the test set. Using that value, we can simply adapt the linear dependence of the full set of data, which results in an almost perfect prediction. To be clearer,  the linear regression estimates that the predicted value $p$ is related to the target $\bar p$  through $\bar p= m p$ (blue points in \ref{generalization}). If there is a single input data in the test set such that the target value $\bar p^*$ is known, then the dressed prediction can be written as $\bar p= m p  (p^*/\bar p^*)$, which results in the red points of Fig. \ref{generalization}.}

\section{Conclusions}

In conclusion, we have introduced a novel QELM that can be used to estimate the degree of entanglement of Werner states. This model demonstrates significant promise for practical applications in quantum information processing. A crucial aspect of our investigation was exploring the QELM's resilience in the presence of noise. Our findings highlight its remarkable robustness, suggesting its applicability in real-world scenarios where data may be corrupted by noise. This resilience significantly enhances the versatility of our method, making it suitable for various practical applications. We also consider the role of the dynamical phases of the reservoir by studying the computational capability of the system as a function of the external magnetic field. 
Our findings differ from previous studies in the field of quantum reservoir computing, where the internal memory of the reservoir is often considered crucial for optimal performance. However, our results suggest a distinct scenario. For our specific QELM model, the optimal performance is achieved around the quantum critical point, a region where the system undergoes a significant phase transition. Conversely, the ergodic phase (a different thermodynamic state) does not offer any significant performance advantage.

Compared to other QELM approaches, our model offers a distinct advantage. For each new instance, the reservoir state is reset, allowing the model to achieve its task solely through its internal dynamics. This simplifies the model and potentially improves its efficiency. Finally, our findings further demonstrate the model's suitability for domain generalization. This means the QELM can effectively learn from data in one domain and apply that knowledge to make accurate predictions in a different but related domain. This broadens the potential applications of our model across various tasks within the realm of quantum information processing and quantum computing.

\begin{acknowledgments}
We acknowledge the Spanish State Research Agency, through the Mar\'ia de Maeztu project CEX2021-001164-M funded by \\ MCIU/AEI/10.13039/501100011033, through the COQUSY project PID2022-140506NB-C21 and -C22 funded by \\ MCIU/AEI/10.13039/501100011033, and the QuantERA QNet project PCI2024-153410 funded by \\ MCIU/ AEI/10.13039/501100011033 and cofounded by the European Union; MINECO through the QUANTUM SPAIN project, and EU through the RTRP - NextGenerationEU within the framework of the Digital Spain 2025 Agenda. The CSIC Interdisciplinary Thematic Platform (PTI+) on Quantum Technologies in Spain (QTEP+) is also acknowledged.
 \end{acknowledgments}

 \appendix
\renewcommand{\thefigure}{A\arabic{figure}}

\setcounter{figure}{0}
 \section{Extended output layer}

In the original model, only local $Z$ observables $\left\langle\sigma_i^z\right\rangle$ were considered, which limited the amount of information extracted from the quantum system. 
  To complete the analysis, we now extend the output layer by incorporating two-point correlations, namely, the expectation value of the product of two Pauli matrices, $\left\langle\sigma_i^z \sigma_j^z\right\rangle$, allowing for a more comprehensive treatment of qubit interactions. This extension provides a larger set of observables, capturing not only individual qubit properties but also correlations between pairs of qubits.

\begin{figure}[b]
    \centering
    \includegraphics[width=1.05\linewidth]{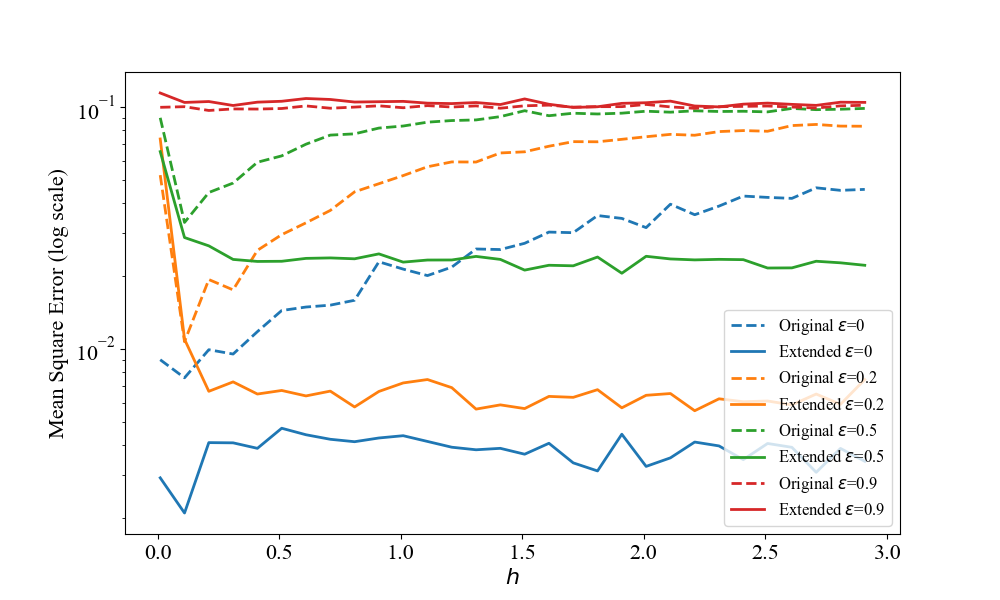}
    \caption{Effect of the extended output layer on {MSE} values under different noise levels ($\varepsilon$) as a function of $h$. The system's parameter are $h=0.1$ and $N=5$ (Average of 20 realizations)}
    \label{fig:extended_av}
\end{figure}

 The effect of this extension is illustrated in Figure \ref{fig:extended_av}, where we compare the correlation performance of the original and extended models across different noise levels $(\varepsilon)$. The dashed lines represent the original model with local $Z$ observables, while the solid lines show the performance of the extended model with two-point correlations. As noise increases $(\varepsilon= 0.2, 0.5, 0.9)$, the extended model consistently demonstrates {lower MSE} values than the original model, highlighting its improved robustness to noise and more detailed measurement capabilities. This enhancement enables the model better to capture the underlying dynamics of the quantum system, resulting in superior performance.

 In many experimental setups, the expected values of the spin-spin correlations can be calculated from the results of projective measurements of individual spins. Indeed, let us assume that we want to compute $\braket{\sigma^z_i}$, $\braket{\sigma^z_j}$ but also $\braket{\sigma^z_i\sigma^z_j}$. For each of the spins, if we run the experiment $N_{m}$ times,  we will have $\braket{\sigma^z_i}=\frac{1}{N_{m}}\sum_{k=1}^{N_{m}} z_{i}^{k}$, where $z_{i}^{k}$ 
 is the $k^{th}$ measurement result.  Then, in order to calculate the expected value of $\braket{\sigma^z_i\sigma^z_j}$, we only need to take the product of the projections and perform the average: $\braket{\sigma^z_i\sigma^z_j}=\frac{1}{M}\sum_{k=1}^{N_{m}} z_{i}^{k} z_{j}^{k}$. So, the extension of the output layer comes with no experimental supplementary cost, even though the cost of the machine learning part is modified.

 \newcommand{\noop}[1]{}
%

%\bibliography{sorsamp}
%\printbibliography

\end{document}